# From Interaction to Collaboration: How Hybrid Intelligence Enhances Chatbot Feedback

JANET RAFNER, Center for Hybrid Intelligence, Department of Management and Aarhus Institute for Advanced Studies Aarhus University, Denmark

RYAN Q GULOY, Adobe Inc.

EDEN W. WEN, Adobe Inc.

CATHERINE M CHIODO, Adobe Inc.

JACOB SHERSON, Center for Hybrid Intelligence, Department of Management, Aarhus University, Denmark

Generative AI (GenAI) chatbots are becoming increasingly integrated into virtual assistant technologies, yet their success hinges on the ability to gather meaningful user feedback to improve interaction quality, system outcomes, and overall user acceptance. Successful chatbot interactions can enable organizations to build long-term relationships with their customers and users, supporting customer loyalty and furthering the organization's goals. This study explores the impact of two distinct narratives and feedback collection mechanisms on user engagement and feedback behavior: a standard AI-focused interaction versus a hybrid intelligence (HI)-framed interaction. Initial findings indicate that while survey measures showed no significant differences in user willingness to leave feedback, use the system, or trust the system, participants exposed to the HI narrative provided more detailed feedback. These findings offer insights into designing effective feedback systems for GenAI virtual assistants, balancing user effort with system improvement potential.

## 1 INTRODUCTION

In HCI, there is a long history of examining how AI systems should be designed. Technological acceptance models have looked at what facilitates users' acceptance of new technological systems [6, 14], and human-centered AI has looked at how to design systems that support human goals and ways of working [21]. A question that has been less-addressed is–to what end? For an organization creating or adopting a new technology, what is the end goal in facilitating adoption?

Authors' addresses: Janet Rafner, janetrafner@mgmt.au.dk, Center for Hybrid Intelligence, Department of Management and Aarhus Institute for Advanced Studies Aarhus University, Aarhus, Denmark; Ryan Q Guloy, Adobe Inc., guloy@adobe.com; Eden W. Wen, Adobe Inc., edenwen@adobe.com; Catherine M Chiodo, Adobe Inc., chiodo@adobe.com; Jacob Sherson, Center for Hybrid Intelligence, Department of Management, Aarhus University, Aarhus, Denmark, sherson@mgmt.au.dk.

Here we explore a fresh approach to this question, proposing that for an organization, the ultimate goal of adopting new technologies such as AI chatbots is to support long-term relationship-building with their customers and users. In this approach, the overarching question is, "what relationship do I aim to build with my users with this technology, and what design elements will enable that relationship?" This question is relevant to organizations of all types, since relationship-building is key to customer retention, attracting donors, empowering employees and many other core organizational goals. For the purposes of this paper we focus on the use case of a for-profit company building software tools used by professionals, and examine the relationship-building potential in an AI chatbot's feedback mechanism.

This study seeks to deepen our understanding of how hybrid intelligence (HI) narratives and features influence user feedback behaviors within AI-driven systems, particularly generative AI (GenAI) chatbots. By situating feedback collection as a co-creative process, HI narratives emphasize collaboration between the system and its users. To investigate this, we implemented two distinct interaction conditions: a conventional AI-focused feedback mechanism and an HI-framed approach, which explicitly highlights the role of user input in shaping and refining the system. The contributions of this work include the design and analysis of an HI narrative framework, which demonstrates that even when traditional survey metrics (e.g., willingness to leave feedback, trust, or system usage) show no significant difference, the elaboration of user feedback can be notably enhanced. These small-scale findings underscore the potential of HI narratives to enrich user engagement and feedback detail, providing actionable insights into how organizations can leverage co-creative mechanisms to build more meaningful, long-term relationships with their users.

## 2 BACKGROUND

### 2.1 GenAI and Chatbots

GenAI refers to computational techniques capable of generating new, context-aware artifacts–such as text, music, and images– distinct from the data they were trained on [4]. GenAI distinguishes itself by its ability to create novel and valuable content, moving beyond the limitations of previous machine learning models [4, 11]. These advancements establish GenAI as a transformative research domain, poised to redefine technological interactions. An increasingly visible application of GenAI lies in its integration with chatbots, automated virtual assistants designed to engage in meaningful conversations. These chatbots have evolved significantly from their origins as rudimentary Q&A systems to sophisticated platforms capable of maintaining social and relational interactions. Leveraging Large Language Models (LLMs) such as OpenAI's GPT, chatbots now exhibit nuanced understanding, emotional intelligence, and contextual adaptability, making them indispensable in customer service, education, and even therapeutic domains [23]. Moreover, the integration of GenAI in chatbots extends their capabilities into co-creative domains. By incorporating real-time feedback loops, these systems can adapt and refine their outputs collaboratively with users, fostering a hybrid intelligence framework where human creativity and machine efficiency synergize [20].

### 2.2 Hybrid Intelligence

Broadly, the concept of Hybrid Intelligence (HI) entails developing optimal synergies, such as co-creation, between humans and algorithms [2, 7, 15]. HI is a computing paradigm and organisational mindset that fosters progressively collaborative human-AI interactions, aiming to establish a partnership between humans and AI rather than treating AI solely as a tool for human augmentation. These systems combine generative power with co-creative interactions to enhance system capabilities. Recent efforts have been made to make HI a holistic framework, incorporating organizational concepts like de- and upskilling from AI deployment [16] and combining interface design considerations with information system and human resource management principles such as business process reengineering and attention to employee psychological safety into 12 concrete design and deployment dimensions [19]. Subsequent work extends this organizational HI perspective to the design of generative AI virtual assistants. This includes an HI-specific technology acceptance model [13, 14], an HI change management framework [19], and design recommendations encouraging AI designers to prioritize psychological safety, user engagement, and continuous innovation [20].

## 2.3 Organizational Partnership with Customers

A common goal for organizations is to improve and deepen their relationships with their customers [5, 10]. Deeper relationships are believed to positively impact trust, loyalty, and customer retention, enabling increased customer lifetime value (CLTV) [3, 22]. Recently, scholars have investigated how chatbots might deepen brands' relationships with their customers by increasing brands' perceived listening [12]. We propose that when we consider organization-public relationships within a computer-mediated interaction, we should conceptualize it as organizational partnership, which includes organizational listening but also includes co-innovation with users, which creates a substantive relationship of two-way communication and organizational responsiveness [8]. In this paper, we propose in-chat feedback mechanisms as a method for organizations to scale this partnership. By collecting user customer feedback within an AI chat experience, we hypothesize that organizations will deepen users' engagement with the chat system, which will contribute to deeper trust and willingness to adopt and propose improvements to the system.

## 2.4 Feedback

One mechanism that has been studied for deepening customer relationships is for companies to engage in social listening [17] which enables bi-directional communication betweens brands and their customers [24]. Bi-directional communication, or interactivity, can deepen brands' relationships with their customers and drive greater brand value [9]. Chatbot interactions give the appearance of bi-directionality, but without a mechanism for collecting and interpreting user feedback, the communication lacks true bi-directionality. As such, we hypothesize that creating a rich and natural feedback experience will support users' sense of bi-directionality, and enhance organization-public relationships.

## 3 STUDY METHODS

### 3.1 Participants

This research effort involved two unmoderated prototype tests conducted with convenience sampled marketing professionals. Participant sourcing and testing were both done on Maze.co. Participant information can be found in 1

Table 1. Participant Characteristics

| Characteristic | Condition 1 Participants | Condition 2 Participants |
| --- | --- | --- |
| Sample Size (N) | 8 participants | 11 participants |
| Job Domain | Marketing | Marketing |
| Job Titles | Marketer (5), Copywriter (1), Director (2) | Marketer (3), Copywriter (1), Director (3), Marketing Manager (1), Brand Manager (1), Entrepreneur (1), Strategist (1) |
| Languages | English speaking | English speaking |

*Age, gender, and highest level of education information are unavailable due to being information that Maze.co does not expose about its participants.*

**Inclusion/Exclusion Criteria:** Participants were required to be marketing professionals familiar with software tools commonly used in their domain. There were no restrictions on years of experience or familiarity with generative AI systems. **Compensation:** Participants were not directly compensated by the researchers; any incentives provided were managed through Maze.co's platform policies.

### 3.2 Study Design

The study employed a between-subjects design with two distinct conditions (see 1), differing in their mode of feedback interaction and the inclusion of explanatory prompts (See appendix for screenshots of the two conditions).

*3.2.1 Conditions.* **Condition 1** (Current condition - Thumbs Up/Down Icons): Feedback was solicited using simple thumbs up/down icons embedded within the AI assistant's responses. This condition provided a streamlined interface, reflecting industry-standard feedback mechanisms. Participants could optionally provide additional comments through a single feedback form.

**Condition 2** (Hybrid Intelligent condition- In-Chat Solicitation with Explanation): Feedback was prompted via an in-chat message that appeared after a qualifying interaction was logged and active use of the AI assistant ceased. This condition included an explanation of how the feedback would be used, aiming to foster a sense of partnership between the user and the AI assistant. Participants were given more space to provide detailed feedback, continuing the conversational tone of the interaction.

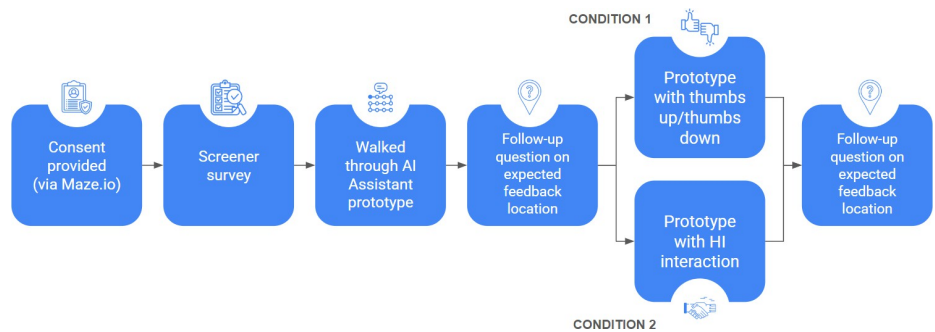

Fig. 1. Process diagram displaying the paths participants took in condition one and two.

*3.2.2 Procedure.* Participants were randomly assigned to one of the two conditions but were not restricted from participating in both. The study spanned from December 23, 2024, to January 27, 2025. Each participant interacted with a prototype AI assistant designed to assist with marketing tasks and then provided feedback based on their experience.

### 3.3 Measures and Analysis

We used both quantitative and qualitative methods to evaluate participant feedback. Quantitative measures included Likert-scale questions assessing willingness to provide feedback, trust in the system, and satisfaction with the interaction. Descriptive statistics summarized survey results, and t-tests for independent samples were applied to compare conditions, with significance thresholds set at $p < 0.05$.

Qualitative feedback was analyzed using a thematic analysis framework, focusing on recurring themes such as ease of use, perceived partnership, and clarity of feedback mechanisms. Open-ended responses were coded into broader categories, such as user experience, feedback intention, and perceived effectiveness, using a schema adapted from [18]'s taxonomy for user feedback classifications.

This mixed-method approach allowed for a comprehensive understanding of both numerical trends and the depth of participant experiences. Efforts were made to ensure consistency in the AI assistant's behavior across conditions, and data from both conditions were compared to evaluate the effectiveness of the HI narrative in eliciting detailed feedback.

| Theme | Codes | Examples |
| --- | --- | --- |
| **Topics** | **Product quality:** feedback on the inherent features and characteristics of the product, such as performance, functionality, and reliability. It assesses what the product is and how well it performs its intended functions. | "Because it needs improvement." (C-C) <br> "It was helpful but could still use tweaks."(C-C) |
| | **Product context:** describes the product's integration with other systems, content accessibility, or general functionality that does not explicitly relate to user experience. | "Pretty much all AI assistants already use this system." (C-C) <br> "I use something similar and would need to see the quality of the work this GPT produces to switch." (C-C) |
| **User Experience** | **Quality in use:** feedback on the user's experience while interacting with the product, focusing on usability, efficiency, task completion, and overall satisfaction. It evaluates how effectively and easily users can achieve their goals with the product. | "It was honestly really straightforward." (HI-C) <br> "It will help perform tasks more quickly." (HI-C) <br> "Speeds up ideation, speeds up execution, AI can suggest elements that I haven't considered." (HI-C) <br> "It was difficult." (C-C) |
| **Intention** | **Informing:** provides information to persuade or dissuade others to use the product, or justifies a rating. | "I'd be more willing, the higher quality answers it gives. Less willing if the system gives answers I would have to interpret a lot." (C-C) |
| | **Reporting:** identifies problems or defects with the product, often directed at the developers. | "My experience was incomplete. The system froze or didn't let me click." (HI-C) |
| | **Requesting:** suggests new features or changes to existing functionality, or expresses a desire for improvements. | "I want there to be a text area for me to type things I didn't like, though I know I can just click 'other'." (HI-C) <br> "I wish it actually said what each thumb meant." (C-C) |

Table 2. Codebook of Analysis Themes, Codes, and Examples

## 4 RESULTS

### 4.1 Survey Measures

Survey results revealed no significant differences between the Hybrid Intelligence (HI) and Current conditions in participants' willingness to leave feedback, use the system, or trust the system. This is not surprising given the small-scale of the study $N_1 = 8; N_2 = 11$.

### 4.2 Feedback Usability and Accessibility

Qualitative feedback emphasized the relative ease of finding and using the feedback mechanism in the HI condition compared to the Current condition. In the Current condition, participants frequently reported difficulties locating the thumbs-up/down buttons, often expressing frustration or confusion:

> *It wasn't that easy; I wouldn't know where to go but I had to guess to press the thumbs up.*
>
> (Copywriter, C-C)

> *A little confusing as it should be at the bottom of the chat.*
>
> (Marketer, C-C)

In contrast, the HI condition's design leveraged a more prominent feedback interface, occupying greater visual real estate, which participants found easier to locate and interpret:

> *There is a very visible box that asks the user if the content is what was expected.*
>
> (Marketer, HI-C)

> *It works well right under; however, I wish it was clearly labeled as seeking feedback.*
>
> (Marketer, HI-C)

Both conditions indicated room for improvement in labeling and placement of feedback buttons, with participants recommending clearer indicators to facilitate interaction.

### 4.3 Feedback Content and Quality

Despite the likert scale questions showing minimal major differences between conditions, the length of written responses was higher in the second condition, as seen in Figure 2.

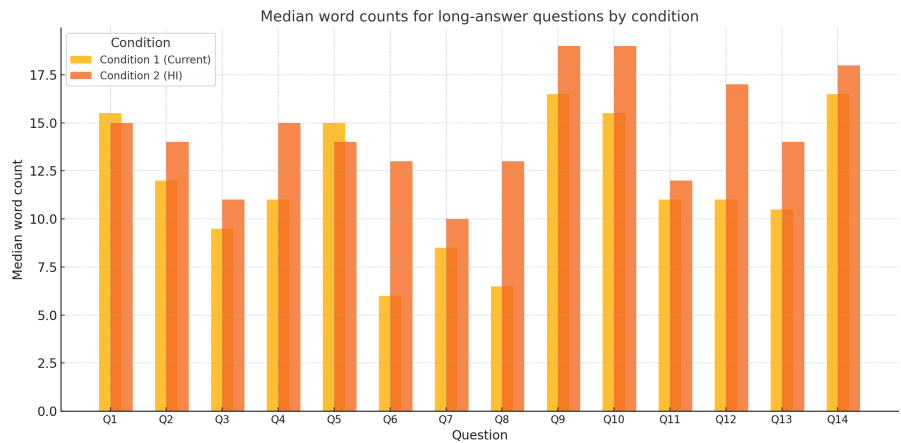

Fig. 2

The analysis of long-answer questions aimed to determine whether participants in the HI condition provided longer and more detailed responses compared to those in the Current condition across open-ended questions. Analysis of long-answer responses revealed an average word count of 15.9 in the HI condition versus 12.0 in the Current condition, a statistically significant difference $t = -3.9$, $p = 0.0001$. These findings indicate that participants in the HI condition provided significantly longer responses on average, suggesting greater engagement or willingness to elaborate in their feedback. This highlights the potential of the HI system to encourage deeper and more reflective input compared to the Current condition.

### 4.4 Coding and Thematic Analysis

For a first exploratory analysis of the content of the qualitative responses we chose three free text questions of particular relevance to concrete interaction design feedback:

(1) **How was your experience in locating and understanding where the feedback system is?**
(2) **How would you rate the feedback system you just used and why?**
(3) **What makes you more or less willing to use an AI system like this as part of your work?**

Among these, User Experience was the most frequently coded theme in both conditions, reflecting participants' focus on usability, ease of navigation, and satisfaction with task-related functionality. The HI condition consistently highlighted seamless feedback mechanisms and intuitive interactions, exemplified by comments such as *"It was honestly really straightforward."* Conversely, participants in the Current condition often reported challenges related to navigation and unclear interface elements, as noted in remarks like *"I had to guess where to go; it wasn't obvious."*

Intention, capturing feedback motivations and suggestions for system improvements, was more prominent in the HI condition (39%) than in the Current condition (30%). Participants in the HI condition frequently requested enhancements or expressed collaborative motivations, indicating deeper engagement with the system. Topics, which included feedback on product quality and context, appeared less frequently overall but were more common in the Current condition (21%) than in the HI condition (10%). Feedback in the Current condition often emphasized broader product-related aspects, such as system integration and functionality, with one participant stating, *"I'd need to see how this integrates with other tools I use daily."*

## 5 DISCUSSION, LIMITATIONS, AND FUTURE DIRECTIONS

Although the sample size precludes statistical significance across all analysis modes, the data suggest meaningful trends. Participants in the Hybrid Intelligence condition provided richer, more intention-driven feedback, likely reflecting greater engagement with the system features. In contrast, participants in the Current condition often emphasized broader product-related aspects, such as system integration and general functionality. These patterns indicate potential areas for further exploration, particularly in refining feedback mechanisms to enhance user experience and address contextual concerns about applicability.

A key finding from our study was the significant increase in the length of feedback provided in the HI condition compared to the Current condition. This aligns closely with the exploratory results of Sherson and Vinchon (2024), who observed a trend toward higher feedback word count in a controlled hybrid intelligence feedback study. In their research, Group 2 participants, who engaged in comparative feedback tasks, provided longer responses (mean of 26 words) than Group 1 participants, who engaged in single-output feedback tasks (mean of 21 words)[20]. Sherson and Vinchon suggest that higher word counts may indicate more deliberate engagement in the evaluation process, with participants offering more justification and detail in their feedback. In the context of our study, the increased length of feedback in the HI condition supports the notion that presenting feedback collection as a co-creative process fosters greater user investment. This finding further reinforces the hypothesis that word count can serve as a proxy for feedback quality, as it reflects the depth of participant engagement and the effort invested in justifying their perspectives [1]. These results offer valuable insights into how co-creative feedback mechanisms might enhance not only the quality of feedback but also user satisfaction and their sense of partnership with the system. While our study was not designed to examine causality, the descriptive patterns observed suggest that hybrid intelligence feedback mechanisms may hold promise for organizations seeking to strengthen user relationships and obtain actionable insights. Future research could build on these findings by exploring the psychological factors that contribute to longer, more detailed feedback and testing additional design features that further amplify user engagement. These findings also emphasize the importance of designing systems that encourage reflective and detailed user feedback. By refining hybrid intelligence mechanisms to elicit specific themes or actionable insights, organizations could unlock the full potential of user input to drive iterative improvements in generative AI systems. Expanding on these exploratory results through larger-scale studies will be crucial for understanding the mechanisms underlying these trends and validating their broader applicability.

Several limitations affected the study's outcomes. The relatively small sample size (11 participants in the HI condition and 8 in the Current condition) limited statistical power, reducing the ability to detect small or medium effect sizes. Additionally, feedback coding relied on participant-generated content, leading to variability in the quantity and richness of responses, which may have influenced coding distributions. Finally, subjective interpretation during qualitative coding may have introduced bias, despite efforts to apply systematic methods.

To address the limitations of this study and build on its findings, future research should prioritize larger-scale studies to increase statistical power and detect meaningful differences in the distribution of feedback themes. Expanding the sample size would allow for a more robust analysis and provide stronger evidence for observed trends. Additionally, future work should focus on designing feedback systems that elicit actionable input aligned with specific organizational goals. The thematic analysis revealed differences in the types of feedback introduced by users in the two conditions, suggesting that targeted system features could help organizations gather feedback with greater operational value.

Moreover, integrating active listening mechanisms into feedback systems presents another promising direction. Systems that summarize user input and solicit additional details could further foster engagement and a sense of partnership between users, the system, and the organization. Exploring these elements could provide deeper insights into how feedback mechanisms can be optimized to strengthen user relationships and improve system outcomes.

## 6 CONCLUSION

While both feedback systems enabled user input, the hybrid intelligence approach led to richer and more detailed feedback. This suggests that framing feedback as a co-creative process can elevate engagement and strengthen the relationship between users and generative AI systems. Expanding on these insights could drive more meaningful improvements in AI design and user experience.

## REFERENCES

[1] Philip C Abrami, Robert M Bernard, Evgueni Borokhovski, Anne Wade, Michael A Surkes, Rana Tamim, and Dai Zhang. 2008. Instructional interventions affecting critical thinking skills and dispositions: A stage 1 meta-analysis. *Review of educational research* 78, 4 (2008), 1102–1134.
[2] Zeynep Akata, Dan Balliet, Maarten De Rijke, Frank Dignum, Virginia Dignum, Guszti Eiben, Antske Fokkens, Davide Grossi, Koen Hindriks, Holger Hoos, et al. 2020. A research agenda for hybrid intelligence: augmenting human intellect with collaborative, adaptive, responsible, and explainable artificial intelligence. *Computer* 53, 8 (2020), 18–28.
[3] Eklou R Amendah, Amarpreet Kohli, Nihar Kumthekar, and Gurmeet Singh. 2023. Impact of financial and nonfinancial constructs on customer lifetime value (CLV): US retailer's perspective. *Journal of Relationship Marketing* 22, 3 (2023), 202–237.
[4] A. Bandi, P. V. S. R. Adapa, and Y. E. V. P. Kuchi. 2023. The power of generative AI: A review of requirements, models, input–output formats, evaluation metrics, and challenges. *Future Internet* 15, 8 (2023), 260. https://doi.org/InsertDOIhereifavailable
[5] Jane L. Bowden. 2009. The Process of Customer Engagement: A Conceptual Framework. *Journal of Marketing Theory and Practice* 17 (2009), 63–74.
[6] Fred D Davis, RP Bagozzi, and PR Warshaw. 1989. Technology acceptance model. *J Manag Sci* 35, 8 (1989), 982–1003.
[7] Dominik Dellermann, Philipp Ebel, Matthias Söllner, and Jan Marco Leimeister. 2019. Hybrid intelligence. *Business & Information Systems Engineering* 61, 5 (2019), 637–643.
[8] Kevin C Desouza, Yukika Awazu, Sanjeev Jha, Caroline Dombrowski, Sridhar Papagari, Peter Baloh, and Jeffrey Y Kim. 2008. Customer-driven innovation. *Research-Technology Management* 51, 3 (2008), 35–44.
[9] Tom Duncan and Sandra E Moriarty. 1998. A communication-based marketing model for managing relationships. *Journal of marketing* 62, 2 (1998), 1–13.
[10] Susan Fournier. 1998. Consumers and Their Brands: Developing Relationship Theory in Consumer Research. *Journal of Consumer Research* 24 (1998), 343–373.
[11] Ian Goodfellow, Yoshua Bengio, and Aaron Courville. 2016. *Deep Learning*. MIT Press. https://www.deeplearningbook.org
[12] Lincoln Lu, Casey McDonald, Tom Kelleher, Susanna Lee, Yoo Jin Chung, Sophia Mueller, Marc Vielledent, and Cen April Yue. 2022. Measuring consumer-perceived humanness of online organizational agents. *Computers in Human Behavior* 128 (2022), 107092. https://doi.org/10.1016/j.chb.2021.107092
[13] Yaoli Mao, Janet Rafner, Yi Wang, and Jacob Sherson. 2023. A hybrid intelligence approach to training generative design assistants: partnership between human experts and AI enhanced co-creative tools. In *HHAI 2023: Augmenting Human Intellect*. IOS Press, 108–123.
[14] Yaoli Mao, Janet Rafner, Yi Wang, and Jacob Sherson. 2024. HI-TAM, a hybrid intelligence framework for training and adoption of generative design assistants. *Frontiers in Computer Science* 6 (2024), 1460381.
[15] Nikhil Prakash and Kory W Mathewson. 2020. Conceptualization and Framework of Hybrid Intelligence Systems. *arXiv preprint arXiv:2012.06161* (2020).
[16] Janet Rafner, Dominik Dellermann, Arthur Hjorth, Dóra Verasztó, Constance Kampf, Wendy Mackay, and Jacob Sherson. 2022. Deskilling, upskilling, and reskilling: a case for hybrid intelligence. *Morals & Machines* 1, 2 (2022), 24–39.
[17] Meryl Rosenblatt MBA, Taylor Curran, and Jessica Treiber. 2018. Building Brands through Social Listening. (2018).
[18] Rubens Santos, Eduard C Groen, and Karina Villela. 2019. A Taxonomy for User Feedback Classifications.. In *REFSQ Workshops*, Vol. 2376.
[19] Jacob Sherson, Baptiste Rabecq, Dominik Dellermann, and Janet Rafner. 2023. A Multi-Dimensional Development and Deployment Framework for Hybrid Intelligence. In *HHAI 2023: Augmenting Human Intellect*. IOS Press, 429–432.
[20] Jacob Sherson, Johannes Rafner, and Sibel Büyükgüzel. 2024. Operational criteria of hybrid intelligence for generative AI virtual assistants. In *HHAI 2024: Hybrid Human AI Systems for the Social Good*. IOS Press, 475–477.
[21] Ben Shneiderman. 2020. Human-centered artificial intelligence: Three fresh ideas. *AIS Transactions on Human-Computer Interaction* 12, 3 (2020), 109–124.

[22] Achim Walter, Thilo A Mueller, and Gabriele Helfert. 2000. The impact of satisfaction, trust, and relationship value on commitment: Theoretical considerations and empirical results. In *IMP Conference Proceedings*. Citeseer, 07–09.
[23] Alice Zhou, Wan-Hsiu Sunny Tsai, and Linjuan Rita Men. 2024. Optimizing AI social chatbots for relational outcomes: The effects of profile design, communication strategies, and message framing. *International Journal of Business Communication* (2024). https://doi.org/10.1177/23294884241229223
[24] Yueming Zou. 2015. Social media review: the impact of social on brand-consumer relationships. In *Ideas in Marketing: Finding the New and Polishing the Old: Proceedings of the 2013 Academy of Marketing Science (AMS) Annual Conference*. Springer, 624–624.

# A SCREENSHOTS FROM THE TWO CONDITIONS

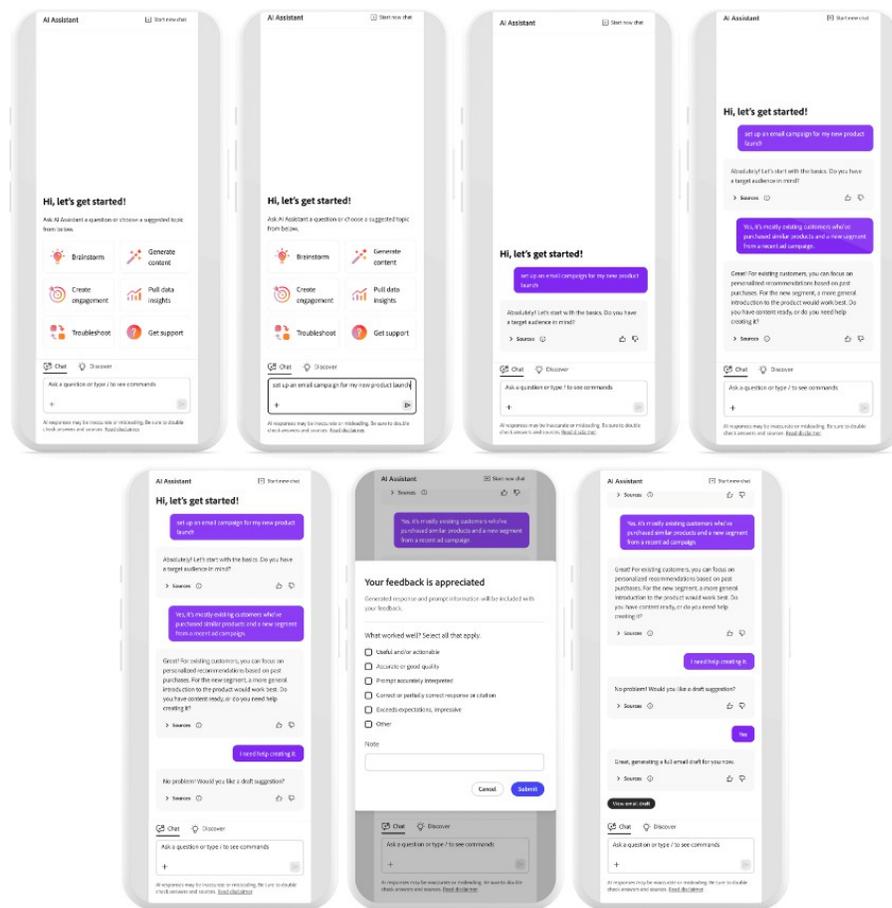

Fig. 3. Screenshot of condition 1 (Current Condition)

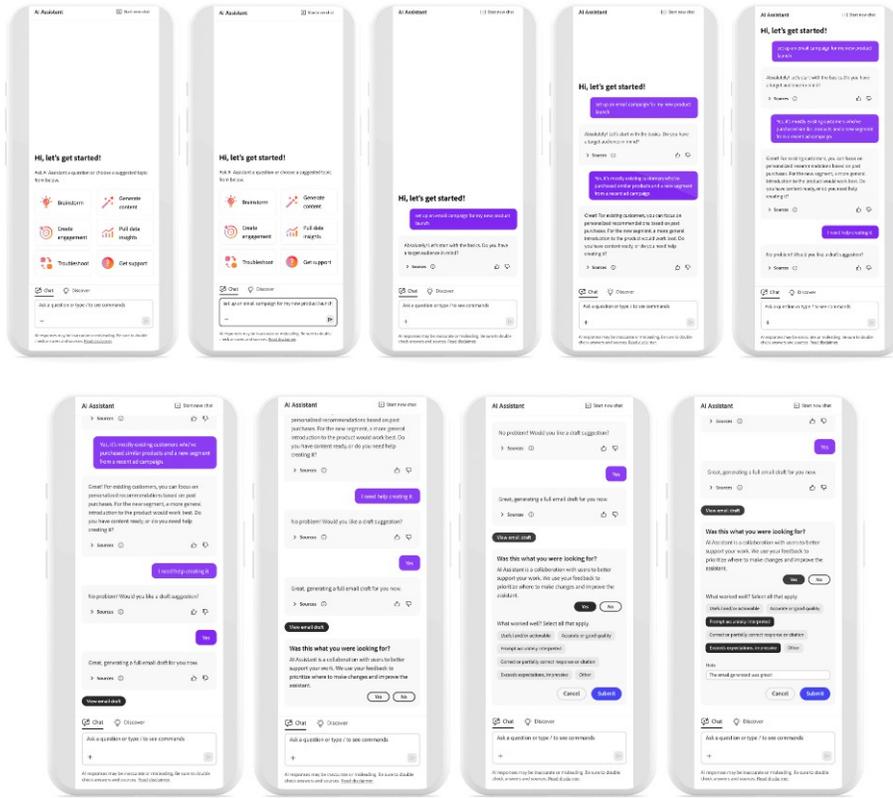

Fig. 4. Screenshot of condition 2 (HI Condition)

B

**SURVEY RESPONSES**

Charts displaying the survey responses from the participants.

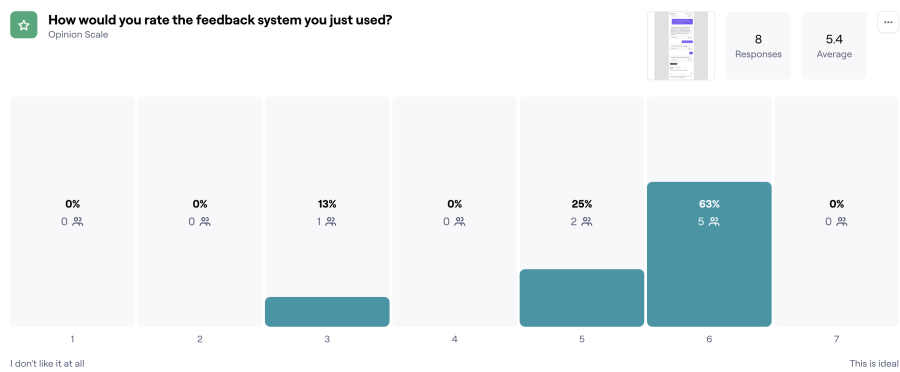

Fig. 5

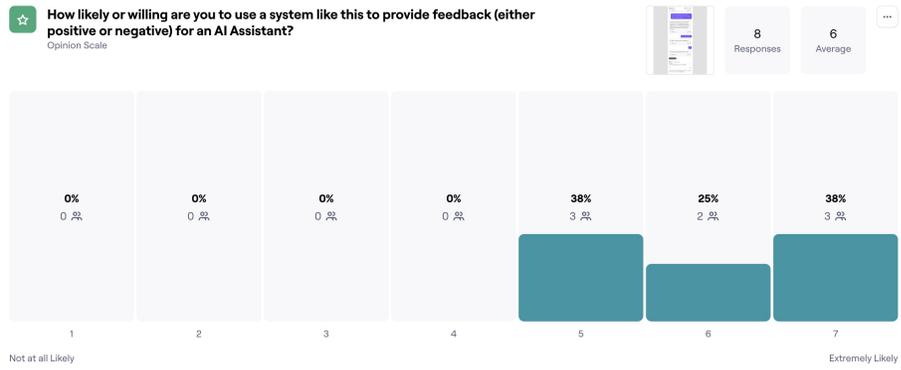

Fig. 6

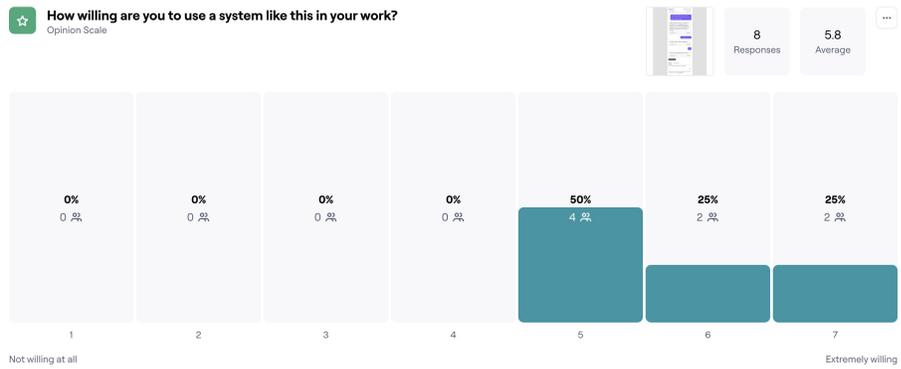

Fig. 7

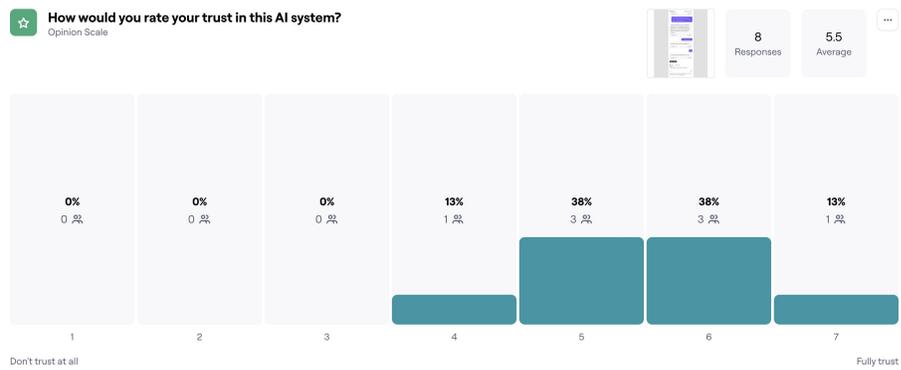

Fig. 8

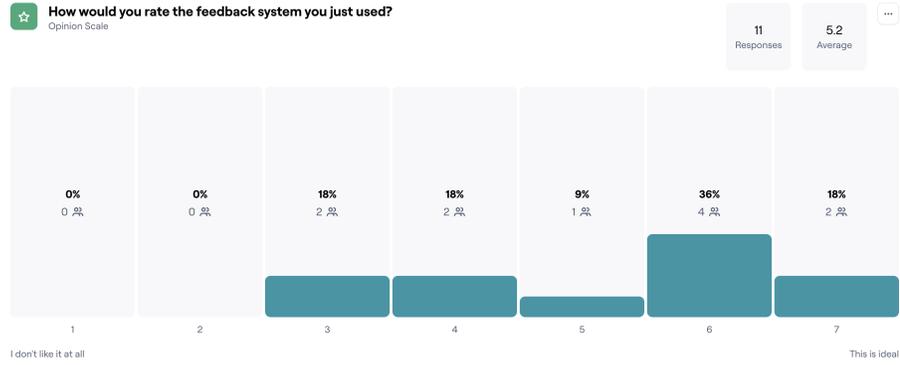

Fig. 9

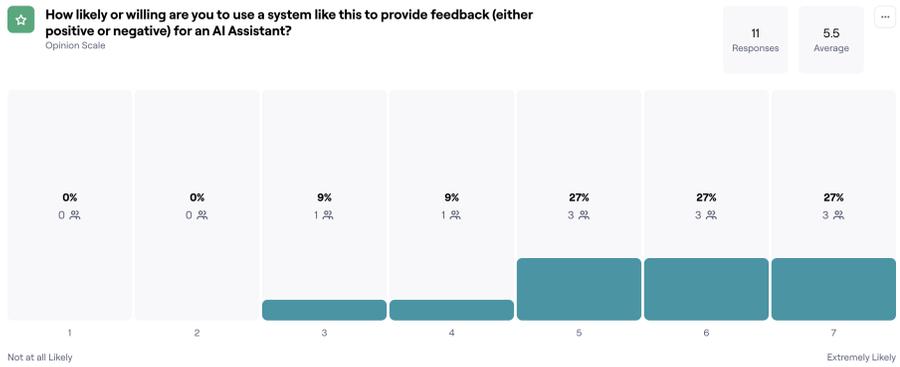

Fig. 10

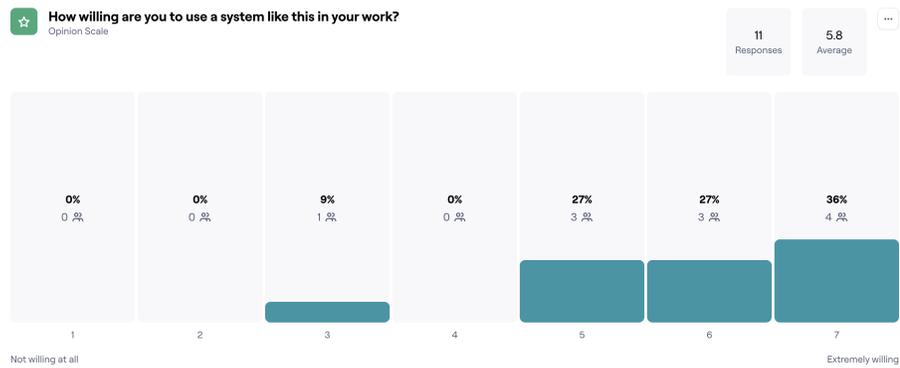

Fig. 11

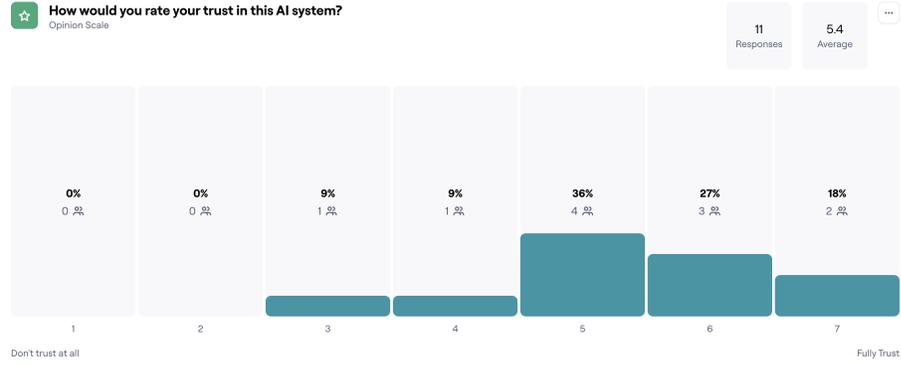

Fig. 12